\pdfoutput=1 

\documentclass[english,twocolumn,showkeys,nofootinbib]{revtex4-1} 
\usepackage[T1]{fontenc}
\usepackage{graphicx}
\usepackage{amssymb,amsmath}
\usepackage{subfigure}
\usepackage{color}
\usepackage{units}
\usepackage{comment}
\usepackage{multirow}
\usepackage{makecell}

\newcommand{\dif}{\mathrm{d}}
\definecolor{grey}{gray}{.35} 
\definecolor{green}{rgb}{0.1, 0.5, 0.1}

\begin{document}

\title{Unfair and Anomalous Evolutionary Dynamics from Fluctuating Payoffs}

\author{Frank Stollmeier}
\affiliation{Network Dynamics, Max Planck Institute for Dynamics and Self-Organization (MPIDS), Am Fa{\ss}berg 17, 37077 G\"ottingen, Germany\\
Institute for Nonlinear Dynamics, Faculty of Physics, University of G\"ottingen, Am Fa{\ss}berg 17, 37077 G\"ottingen, Germany}
\author{Jan Nagler}
\email[Corresponding author. E-mail: ]{jnagler@ethz.ch}
\affiliation{Computational Physics for Engineering Materials, IfB, ETH Zurich, Wolfgang-Pauli-Strasse 27, CH 8093 Zurich, Switzerland}

\keywords{evolutionarily stable state; payoff fluctuations; ergodicity breaking}

\begin{abstract}
Evolution occurs in populations of reproducing individuals.
Reproduction depends on the payoff a strategy receives.
The payoff depends on the environment that may change over time, 
on intrinsic uncertainties, and on other sources of randomness. 
These temporal variations in the payoffs can affect which traits evolve.
Understanding evolutionary game dynamics that are affected by varying payoffs remains difficult. 
Here we study the impact of arbitrary amplitudes and covariances of temporally varying payoffs on the dynamics.
The evolutionary dynamics may be ''unfair``, meaning that, on average, two coexisting strategies may persistently receive different payoffs.
This mechanism can induce an anomalous coexistence of cooperators and defectors in the Prisoner's
Dilemma, and an unexpected  selection reversal in the Hawk-Dove game.
\end{abstract}

\maketitle

How species interact 
depends on the environment and is thus often uncertain or subject to ongoing variations.
Traditional game theory has assumed constant payoff structures.
Here, we demonstrate by independent methods that the
dynamics of averaged payoff values does not well approximate the dynamics of fluctuating payoff values. 
We show that payoff fluctuations  induce qualitative changes in the dynamics.
For instance, a Prisoner's Dilemma with payoff fluctuations may have the evolutionary dynamics of a Hawk-Dove game with constant payoff values. As a consequence, cooperators can coexist with defectors -- 
without any further cooperation maintaining mechanism 
such as kin or  group selection \cite{Traulsen2006b,Lehmann2007},
reciprocity \cite{Nowak2006five}, or spatial structures \cite{Nowak1992}. \\
First of all, how environmental fluctuations and payoff stochasticities affect the evolution of interacting species depends on the time scales. 
If the fluctuations are much faster than reproduction, 
adaptation reaches a stationary state where species are adapted to living in a rapidly fluctuating environment. 
If the fluctuations are much slower than the generation time (e.\,g. ice ages or geomagnetic field reversals),  
adaptation quickly reaches a stationary state 
which slowly drifts to follow the fluctuation.
Ultimately challenging is the case when the fluctuations and reproduction are at a similar pace such that 
adaptation is continuously following the environmental changes. 
Here, we show that such states are subject to noise-induced transitions.
Noise-induced transitions have been studied in dynamical systems, 
where the most prominent models study the effects of additive noise \cite{Broeck1997,Toral2011,Horsthemke1984}.
In dynamical systems, both additive and multiplicative noise can lead to an array of anomalous noise-induced effects such as stochastic resonance \cite{Gammaitoni1998} and the creation of stable states \cite{Lipshtat2006,Biancalani2015}.
We wish to investigate the consequences of multiplicative noise in evolutionary game theory that have not been systematically studied yet.

A number of studies used stochastic models of population extinction to analyze the impact of environmental stochasticity 
on the extinction risk of small and large populations \cite{Leigh1981,Lande1993,Foley1994}.
Particular attention has been spent on how the species' mean time to extinction depends on
a small randomly varying growth rate \cite{Ovaskainen2010}, and on the autocorrelation 
of the environmental noise \cite{Schreiber2010,Morales1999,Heino2000,Wilmers2007,Schwager2006,Heino2003,Ruokolainen2009,Greenman2005,Kamenev2008}. 
Likewise in evolutionary game theory, the question of how 
fixation, i.\,e.\;the transition to the survival of only one species,
depends on environmental stochasticity attracted a lot of attention \cite{Nowak2004,Traulsen2006,Altrock2009,Assaf2013,Ashcroft2014,Houchmandzadeh2015}.
Recently, how the fixation depends on environmental stochasticity was also studied in the case of multi-player games \cite{Baron2016}.

As opposed to these efforts, we will focus on the impact 
of payoff fluctuations 
on the stationary states.\\
Environmental fluctuations have been integrated in models for evolutionary games in different ways, 
including fluctuating reproduction rates \cite{Foster1990, Fudenberg1992, Hofbauer2009, Traulsen2004}, selection strength \cite{Assaf2013} and population size \cite{Houchmandzadeh2012, Houchmandzadeh2015,Huang2015,Gokhale2016,Constable2016}. 
We integrate environmental fluctuations as varying payoff values to study 
 situations in which the environmental fluctuations affect the way the species interact. 
Thereby we assume that all individuals experience the same environment, meaning that the payoff values vary with time but not between individuals.

We explore the landscape of dynamical changes of evolutionary games induced by such fluctuating payoffs.  
We consider both deterministic (e.\,g.\;seasonal) as well as stochastic fluctuations with varying intensities and correlations. 
For a realistic description it is necessary to also include intrinsic noise in finite populations \cite{Lande1993,Taylor2004,Nowak2004,Traulsen2006}.
However, we aim to reveal phenomena that were unknown so far because they were hidden by the idealized assumption of constant payoffs.
Therefore we isolate the effects of fluctuating payoffs from the diverse effects of intrinsic noise in finite populations by 
studying the replicator equation, which 
describes the evolution of strategies in infinite populations, and 
the Moran process \cite{Moran1958random} for finite but large populations.

\paragraph*{Anomalous evolutionarily stable states}

\begin{figure}
    \centering
    \includegraphics[]{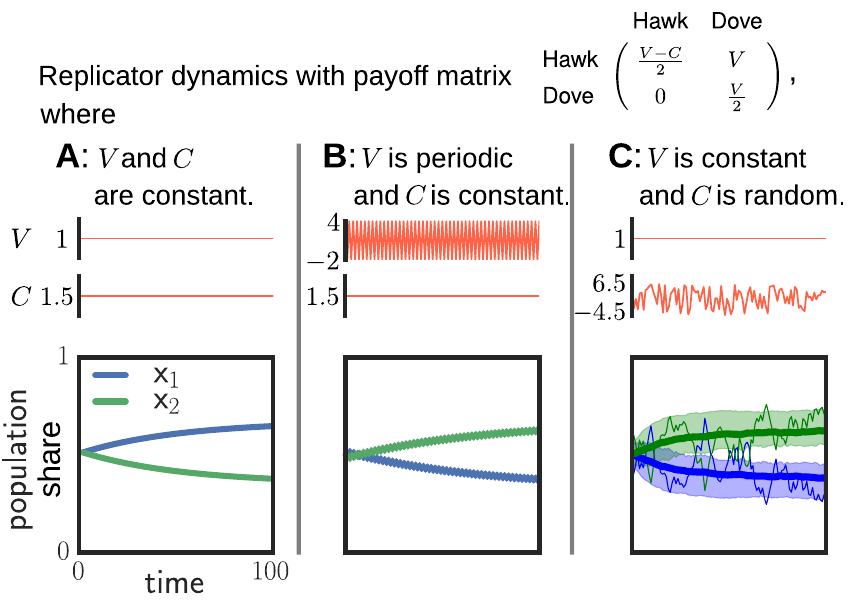}
    \caption{    {\bf
    Selection reversal
    in a Hawk-Dove game with constant, periodic and random payoff.} 
    (A) describes a traditional Hawk-Dove
    game. The population starts at $x_1=x_2=0.5$ (50 \% Hawks, 50\% Doves) and converges to an evolutionarily stable state where $x_1>x_2$.
    Periodically (B) or randomly fluctuating payoffs (C)
    shift the evolutionarily stable state such that 
    $x_1<x_2$.
    } 
    \label{fig:simple_example}
\end{figure}

\begin{figure}
    \centering
    \includegraphics[]{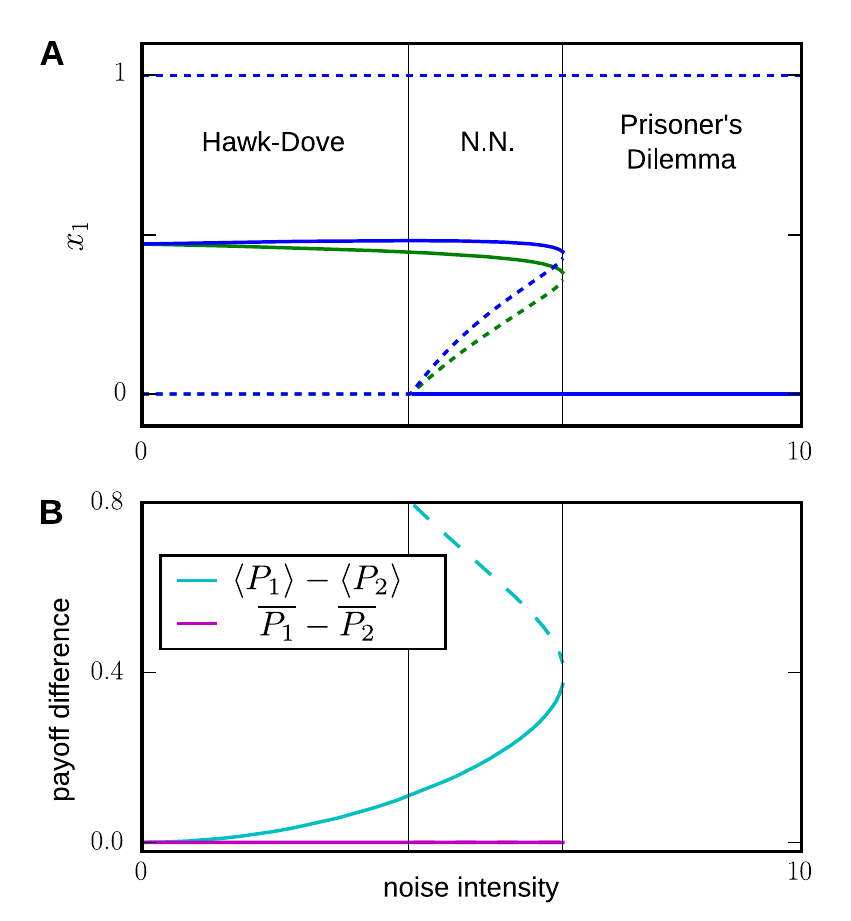}
    \caption{
    {\bf Fluctuations transform a Hawk-Dove game into a Prisoner's Dilemma and cause ``unfair'' stable coexistence.}
    (A) Shown is the anomalous stationary state (solid line: stable, dashed line: unstable) 
    of the fraction of cooperators $x_1$ as a function of the noise intensity. Due to alternating payoff values the stationary states consist of two periodic points (green and blue).
    With increasing intensity, the dynamical structure of a Hawk-Dove game first changes to a game without analog in traditional games (N.N.) and finally to a Prisoner's Dilemma game.
    (B) The difference of the averaged payoffs received by the two players corresponding to the stationary states of coexistence in A.
    In the arithmetic mean the received payoffs are unfair. 
    In the geometric mean they are equal, as predicted by Eq.~(\ref{eq:geometric_mean_payoff}). 
} 
    \label{fig:transition}
\end{figure}

Multiplicative growth is a common model that underlies both population and evolutionary dynamics.
In the simple case of time-discrete exponential growth, the population number $n$ is described by
$ n_{t+1} = r n_t$. Depending on the growth rate $r$, the population will diverge ($r>1$), remain constant ($r=1$) or decay ($0 \le r < 1$).
However, a time-dependent growth rate $r_t$ can lead to intricate results. 
As an example, compare a growth rate that is switching between $1$ and $1.1$ with a growth rate that is switching between $0.6$ and $1.5$. Both have the same arithmetic average that is greater than one, but the population will diverge in the first case because $1\cdot 1.1=1.1$ and decay in the second case because $0.6\cdot 1.5=0.9$. In general, the long-term growth is determined by the geometric mean of the growth rate $\bar{r}$, and the population will diverge if $\bar{r}>1$, remain constant if $\bar{r}=1$ and decay if $0 \le \bar{r} < 1$. Like in this example, multiplicative noise has generally a net-negative effect on growth in the long-term \cite{Lewontin1969, Peters2011, Peters2013}.\\
Models of evolutionary game theory 
are more complex but
share the same underlying property, which leads to noise-induced non-ergodic behavior.

In the classical Hawk-Dove game two birds meet and compete for a shareable resource $V$, the positive payoff. 
If a Hawk meets a Dove the Hawk alone gets the resource, if two Doves meet they share the resource and if two Hawks meet they fight for the resource, which costs energy and implies the risk of getting injured, formalized by a negative payoff $-C$.
Since 50\% of the Hawks win and 50\% of the Hawks loose a fight,
the average payoff of a Hawk meeting a Hawk in the limit of an infinite population is $\frac{V-0}{2}+\frac{0-C}{2}=\frac{V-C}{2}$. \\
Fig.~\ref{fig:simple_example} (A) shows that for $V=1$ and $C=1.5$ the time-discrete replicator dynamics leads to an evolutionarily stable state in which a larger population of Hawks coexists with a smaller population of Doves. 
However, in a changing environment the payoff matrix will not be constant. For example, the abundance of the food resource may change periodically with the seasons, or the risk of death caused by an injury may depend on the presence of predators. Fig.~\ref{fig:simple_example} (B) and (C) show how the evolutionarily stable state can change if $V$ or $C$ fluctuate such that their averages are still the same as in (A).
Similar to the aforementioned example with the exponential growth process, the noise has a net-negative effect on the long-term growth of the strategies in replicator dynamics, too. Due to the specific structure of the Hawk-Dove game payoff matrix, the negative effect of the noise of both $V$ and $C$ is stronger for the population of Hawks than for the Doves, such that with sufficient noise the Doves dominate the population in the evolutionarily stationary state. 
Next, we show that these anomalous effects are generic for evolutionary games.

In evolutionary game theory the interactions are usually formalized in a payoff function, 
which specifies the reward from the interaction with another player
that is received by a given individual. 
In the simplest case, a game with two strategies is determined 
by a payoff matrix $M$ with $2\times 2$ matrix elements. 
We describe the state of the population as
$\mathbf{x}$ ($\sum{x_i}=1$), where $x_i\ge 0$ is the fraction of players
with strategy $i\in\{1,2\}$. 
Players with strategy $i$ 
receive the payoff $P_i=(M \mathbf{x})_i + b$, 
where the background fitness $b$ ensures that the payoff is positive.
The
assumption that species that receive a higher payoff reproduce faster can be formalized by the replicator equation, 
which is used here in its time-discrete form \cite{Taylor1978}
\begin{align}
&x^{(t+1)}_i =  x_i^{(t)} \cdot r_i(\mathbf{x}^{(t)},M) \label{eq:discrete-replicator-equation}, \\
\text{with } &r_i(\mathbf{x}^{(t)},M)= \frac{(M\mathbf{x}^{(t)})_i+b}{\mathbf{x}^{(t)T} M\mathbf{x}^{(t)}+b} = \frac{P_i}{\langle P \rangle} \label{eq:growth_rate}
\end{align}
and the average payoff of the population $\langle P \rangle = x_1 P_1 + x_2 P_2$. \\ 
Following Smith \cite{Smith1982},
``a population is said to be in an `evolutionarily stable state' [henceforth ESS] if its genetic composition is restored by selection after a disturbance, provided the disturbance is not too large.'' 
Hence the ESS 
describe the long-term behavior of the system
and
are stable stationary states of Eq.~(\ref{eq:discrete-replicator-equation}). 
For a constant payoff matrix $M$, the stationary states $\mathbf{x}^*$ satisfy $r_i(\mathbf{x}^*,M)=1$. 
If two species coexist,
 $r_1(\mathbf{x}^*,M)=r_2(\mathbf{x}^*,M)$ implies that both receive 
 the same payoff $P_1=P_2=\langle P\rangle$, as otherwise
  the species with the higher payoff would 
  move the system away from this state due to faster growth. \\
Now consider continuously changing payoffs with finite means.
The stationary states $\mathbf{x}^*(t)$ are solutions of 
\begin{align}
\overline{r_i(\mathbf{x}^*,M)} := \lim_{T\rightarrow\infty} \left( \prod_{t=0}^{T-1} r_i(\mathbf{x}^*(t),M^{(t)}) \right)^\frac{1}{T} = 1 \text{,} \label{eq:geometric-mean} 
\end{align}
where $M^{(t)}$ is the time-dependent payoff matrix.
Equation~(\ref{eq:geometric-mean}) defines the geometric average, indicated henceforth by the bar.
If the payoff matrix changes deterministically with period $T$ a stationary state is a periodic function $\mathbf{x}^*(t)=\mathbf{x}^*(t+T)$; if it changes randomly a stationary state is a random function $\mathbf{x}^*(t)$ with distribution $\rho^*(\mathbf{x})$.
But how does one calculate the stationary states for periodically and randomly changing payoff matrices?
In contrast to normal ESS 
the stationary states are not solutions of $\langle P_1 \rangle = \langle P_2\rangle$,
where $\langle P_i \rangle := \lim_{T\rightarrow\infty} \frac{1}{T}\sum_{t=0}^{T-1} \left( M^{(t)} \mathbf{x}^*(t)\right)_i$ is the arithmetic time average of the received payoff.

Equation~(\ref{eq:geometric-mean}) implies that
${\overline{r_1(\mathbf{x}^*,M)} = \overline{r_2(\mathbf{x}^*,M)}}=1$, 
and, using Eq.~(\ref{eq:growth_rate}), that 
\begin{align}
\overline{ P_1 } = \overline{ P_2 }. \label{eq:geometric_mean_payoff}
\end{align}
If the fluctuations are small, we can approximate the geometric mean by $\overline{P_i} = \langle P_i\rangle - \frac{\sigma_i^2}{2\langle P_i\rangle } + \mathcal{O}(\sigma_i^4)$ (see Supplementary Material S1), 
where $\sigma_i^2=\text{Var}[P_i]$. 
Using this approximation in Eq.~(\ref{eq:geometric_mean_payoff}) yields 
\begin{align}
\langle P_1 \rangle - \frac{\sigma_1^2}{2\langle P_1 \rangle} = \langle P_2 \rangle - \frac{\sigma_2^2}{2\langle P_2 \rangle} \label{eq:approx_average_payoff}
\end{align}
Equation~(\ref{eq:approx_average_payoff}) shows that $\langle P_1 \rangle$ and $\langle P_2\rangle$ are generally different, which is why we call these stationary states unfair. It includes the case of constant payoff values as a special case\footnote{Note that $\sigma_1$ and $\sigma_2$ depend on the stationary state $x_1$ and the variance and covariance of the payoff values $M=[m_1,m_2,m_3,m_4]$. 
If $\sigma_1=\sigma_2=0$, Eq.~(\ref{eq:approx_average_payoff}) reduces to $\langle P_1 \rangle = \langle P_2 \rangle$. 
For small fluctuations we can approximate them as 
$ \sigma_1^2 \approx \text{E}[x_1]^2 \text{Var}[m_1] + (1-\text{E}[x_1])^2 \text{Var}[m_2] + 2(\text{E}[x_1]-\text{E}[x_1]^2)\text{Cov}[m_1,m_2]$
and
$\sigma_2^2 \approx \text{E}[x_1]^2 \text{Var}[m_3] + (1-\text{E}[x_1])^2 \text{Var}[m_4] + 2(\text{E}[x_1]-\text{E}[x_1]^2)\text{Cov}[m_3,m_4]$.}.
Figure \ref{fig:transition} (A) illustrates how 
payoff fluctuations may change the evolutionary dynamics and thereby transform one game into another game.
Figure \ref{fig:transition} (B) shows how the arithmetic and the geometric average of the payoffs the two species receive deviate
(see also Supplementary Fig.~S1).

\paragraph*{Deterministic payoff fluctuations}

\begin{figure}
    \centering
    \includegraphics[]{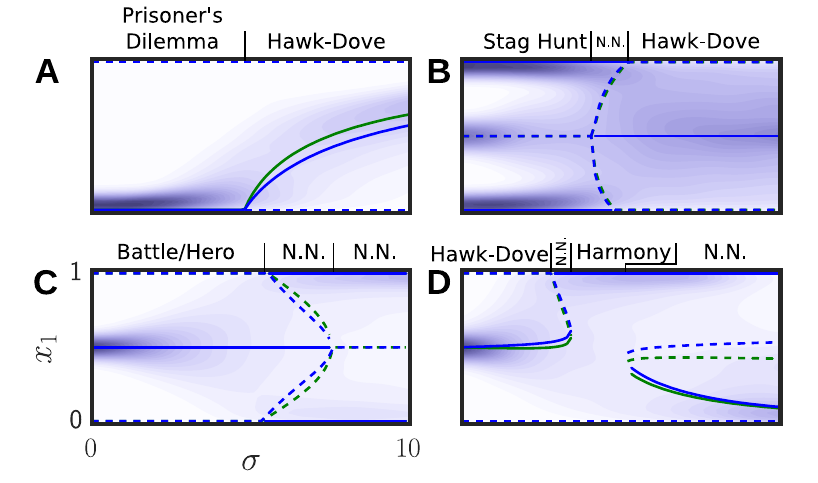}
    \caption[please allow matrices cause we do not need them really]{{\bf  Evolutionarily stable states with increasing fluctuation intensity.}
    Stable and unstable states (solid and dashed lines) $x_1^*(\sigma)$ for  
    games with alternating payoff fluctuations (blue and green are the two periodic points). 
    The payoff matrices are 
                      $M^{(t)}=[ 3, 1, 4, 2]+(-1)^t \sigma [ 0 , 0, 0, 1]$ in (A),
                      $M^{(t)}= [4,1,3,2] +(-1)^t \sigma [ 1, 0, 0 ,1]$ in (B),
                      $M^{(t)}=[ 2, 3, 4, 1]+(-1)^t \sigma [0 , 1.3, 1.3 , 0]$ in (C)
                  and $M^{(t)}=[ 3, 2, 4, 1] +(-1)^t \sigma [ -0.75 , 1, -2 , 1]$ in (D). 
                      In each example the background fitness is $b=10$. The names of the games are identified using criteria described in the Supplementary Material S3. \\ 
                      For the same games but stochastic instead of alternating noise, the background shows the average of three stationary distributions resulting from the initial distributions $\delta(x)$, $\delta(x-0.5)$ and $\delta(x-1)$.}
    \label{fig:replicator-bifurcations}
\end{figure}

We first consider deterministic payoff fluctuations under the replicator equation (Eq.~(\ref{eq:discrete-replicator-equation})). 
To find the stationary state $\mathbf{x}^*$ we solve Eq.~(\ref{eq:geometric-mean}).
We assume that $M^{(t)}$ is a sequence with period $T$. Consequently, the stationary state $\mathbf{x}^{*(t)}$
is periodic as well and $P(\mathbf{x},M)=\frac{1}{T}\sum_{t=0}^T \delta(\mathbf{x}-\mathbf{x}^{*(t)}) \delta(M-M^{(t)})$. 
Equation~(\ref{eq:geometric-mean}) 
reduces to
\begin{align}
 \overline{r_i(\mathbf{x}^*,M)} = \left( \prod_{t=t'}^{t'+T} r_i\left(\mathbf{x}^{*(t)},M^{(t)}\right)\right)^\frac{1}{T} &= 1.  \label{eq:geometric-mean-discret}
\end{align}
Note that Eq.~(\ref{eq:geometric-mean-discret}) 
has only one free variable because if one periodic point $\mathbf{x}^{*(t')}$ is given, the others are determined by Eq.~(\ref{eq:discrete-replicator-equation}). \\ 
As an illustrative  example, 
 assume an alternating payoff matrix $M^{(t)}=\overline{M}+(-1)^t \sigma\tilde{M}$. 
Then $\mathbf{x}^{*(t)} = \overline{\mathbf{x}}^* + (-1)^t \Delta \mathbf{x}^* $ has the same form and can be found by solving 
Eq.~(\ref{eq:geometric-mean-discret}), which reduces to
\begin{align}
\overline{r_i(\mathbf{x}^*,M)} = \sqrt{ r_i(\mathbf{x}^{*(t)}, M^{(t)})\cdot r_i(\mathbf{x}^{*(t+1)}, M^{(t+1)})} = 1. \label{eq:geometric-mean-simple}
\end{align}
Figure \ref{fig:transition} shows 
the stationary states of a game with the payoff function
\begin{align}
M^{(t)}=\begin{pmatrix} 1.1 & 0.8\\ 2 & 0 \end{pmatrix}+(-1)^t \sigma \begin{pmatrix} -0.33 & 1\\ 1 & 0\end{pmatrix}
\end{align}
For $\sigma=0$ this is a Hawk-Dove game. 
For small $\sigma$, in fact, the stationary states 
predicted by Eq.~(\ref{eq:geometric-mean-simple}) slightly 
deviate from the ESS of the Hawk-Dove game. 
There is a first bifurcation at $\sigma\approx 4.07$, 
from one stable stationary state (solid curves) to two.
At $\sigma\approx 6.4$
there is a second 
bifurcation where the first branch, the stable coexistence, disappears.
The bifurcation behavior induces a pronounced hysteresis effect. %
Ergodicity breaking causes anomalous 
player's payoff expectations  
 as shown
 in Fig.~\ref{fig:transition} (B).
The arithmetic mean of the payoff difference that the players receive
also shows a pronounced hysteresis effect. 
For the geometric mean, as predicted by Eq.~(\ref{eq:geometric_mean_payoff}),
this effect is absent. \\
More generally, fluctuations can even change the number, the positions and the stability of stationary states 
and the dynamics can be structurally very different from the dynamics of 
games with constant payoffs, as shown in Fig.~\ref{fig:replicator-bifurcations}. 
In Fig.~\ref{fig:replicator-bifurcations} (A) large fluctuations induce the onset of cooperation 
for the Prisoner's dilemma as it is effectively transformed to a Hawk-Dove game with stable coexistence.
Figures~\ref{fig:replicator-bifurcations} (B), (C) and (D) show how increasing fluctuations successively transform three other classical games either into different classical games or into games without classical analogs (denoted at ``N.N.'').

In the Supplementary Material S2, 
we show how anomalous stationary states arise from (correlated) stochastic payoffs,
which is mathematically more involving but shows similar effects as from deterministic fluctuations.

\paragraph*{Discussion}
Payoff noise 
in evolutionary dynamics is 
multiplicative 
 and as such causes ergodicity breaking.
The consequences have intricate effects on the coevolution of strategies. Depending
on the details of the system, on the intensity of the fluctuations and even on their covariance,
ergodicity breaking
leads to shifting the payoffs out of equilibrium, shifting the stationary states 
and thereby to fundamental
structural changes of the dynamics. \\
In evolutionary games with constant payoffs, the condition for stable coexistence is that all species have equal growth rates. 
With fluctuating payoffs this condition generalizes to equal time-averaged growth rates, 
which typically
are different from ensemble averages in non-ergodic systems. 
When one naively replaces fluctuating payoffs with their average values, the ensemble averages of the growth rates are recovered but these averages do not correctly predict the dynamics.

Games with fluctuating payoffs require a novel classification that cannot be based on payoff ranking schemes. 
We developed a classification that
primarily considers the dynamical structure (Supplementary Material S3). 
Our classification for evolutionary games 
may be applied to evolutionary games where the payoff structure cannot be described by a simple payoff matrix,
or when other modifications affect the dynamical structure. 
Examples include complex interactions of microbes such as cooperating and free-riding yeast cells, where 
the payoff is a nonlinear function of the densities \cite{Gore2009}. \\
Payoff fluctuations can cause two strategies that coexist in an evolutionarily stable state to receive different time-averaged payoffs.
However, these ``unfair'' stable states are not mutationally stable.
Mutations, in fact,  would turn the ``unfair'' stable state into a meta-game, where the beneficiary aims to increase and the victim aims to escape the unfairness. 
Strategies of this meta-game could be tuning the
adaptation or reproduction rate according to the environmental fluctuation \cite{Traulsen2004}.
Phenotypic plasticity \cite{Pigliucci2005} and bet-hedging \cite{Bergstrom2014} may reduce the necessity to adapt at all. \\
In general, the understanding of evolutionary games in fluctuating environments may be particularly relevant to understanding and controlling microbiological systems. 
Examples for evolutionary games in microbiology are diverse and include yeast cells \cite{Gore2009}, viruses \cite{Turner2003} and bacteria \cite{Kirkup2004,Griffin2004,Dugatkin2005,Yurtsev2013}. 
Because many of these microbes evolve in natural and artificial environments which are fluctuating,
the presented effects are relevant in biotechnology
and healthcare. 
A stable coexistence of antibiotic-sensitive bacteria with antibiotic-degrading bacteria has been proven to be a stable state of a Hawk-Dove-like game \cite{Dugatkin2005,Yurtsev2013}
if the antibiotic concentration is constantly above the concentration which the sensitive bacteria could tolerate alone.
Our framework qualitatively describes the 
competitive interplay of bacteria strains in a fluctuating environment,
for instance, in a patient who is given a daily dose of antibiotic instead of a continuous infusion.\\
Simple experimental settings can directly demonstrate the consequence of non-ergodic anomalous long-term behaviors
in microbiological systems.
Expected shifts and bifurcations in the stationary states of strategies for two strains, or species, competing for resources (and survival) suggest
to study the (co)evolutionary dynamics for a fluctuating control parameter $c$ that, e.g., 
switches between two levels in a square-wave fashion, $c=[c^+,c^-,c^+,c^-,c^+,\ldots]$, where $c^+=c+A$ and $c^-=c-A$.
For increasing fluctuation amplitude $A$, the stationary state is expected to shift, or to change discontinuously,
both as a result of ergodicity breaking. The strongest effect is expected for fluctuations that are of the same time scale as the reproduction period of the model organisms.
However, quantitative predictions require much more specific model systems \cite{deVos2017}.

To conclude, caution is advised when predictions are based on averaged observables, 
in particular, averaged payoffs structures. Our framework predicts anomalous 
stationary states as a generic result of ergodicity breaking in evolutionary dynamics 
that depend on the amplitude and covariance of the fluctuations.

We thank Christoph Hauert for comments on the manuscript.
F.S. acknowledges funding through the International Max Planck Research School (IMPRS) ``Physics of Biological and Complex Systems''.
J.N. acknowledges support from the ETH Risk Center (grant no.\ RC SP 08-15) and from SNF (grant \textit{The Anatomy of Systemic Financial Risk}, no. 162776).


\clearpage
\onecolumngrid
\appendix

\renewcommand{\thepage}{S\arabic{page}} 
\renewcommand{\thesection}{S\arabic{section}}  
\renewcommand{\thetable}{S\arabic{table}}  
\renewcommand{\thefigure}{S\arabic{figure}}
\renewcommand{\theequation}{S\arabic{equation}}
\setcounter{page}{1}
\setcounter{section}{0}
\setcounter{table}{0}
\setcounter{figure}{0}
\setcounter{equation}{0}

\part*{Supplementary Information}
\vspace{1.5cm}

\section{Approximation of the geometric mean}
\label{sec:approximation-of-geometric-mean}

Let $X$ be a random variable with $\text{E}[X]=\mu$, $\text{E}[(X-\text{E}[X])^2]=\sigma^2$ and $\text{E}[(X-\text{E}[X])^3]=0$. We can write the geometric mean of $X$ as 
\begin{align}
\overline{X} = \overline{\mu + \sigma Y} = \lim_{T\to \infty} \prod_{t=0}^{T} (\mu + \sigma y_t)^{\frac{1}{T}} ,
\end{align}
where $Y$ is a random variable with $\text{E}[Y]=0$, $\text{E}[(Y-\text{E}[Y])^2]=1$ and $\text{E}[(Y-\text{E}[Y])^3]=0$.  
Now we have the geometric mean as a function of $\sigma$ and can write the Taylor series of $\overline{X}(\sigma)$ at $\sigma=0$,
\begin{align}
\overline{X} &= \overline{X}(0) + \frac{\dif\overline{X}}{\dif\sigma}(0) \sigma + \frac{\dif^2\overline{X}}{\dif\sigma^2}(0) \frac{\sigma^2}{2} 
                + \frac{\dif^3\overline{X}}{\dif\sigma^3}(0) \frac{\sigma^3}{6} + \mathcal{O}(\sigma^4) \\
  &= \mu - \frac{\sigma^2}{2\mu} + \mathcal{O}(\sigma^4)
\end{align}

\begin{figure}[htp] 
    \centering
    \includegraphics[]{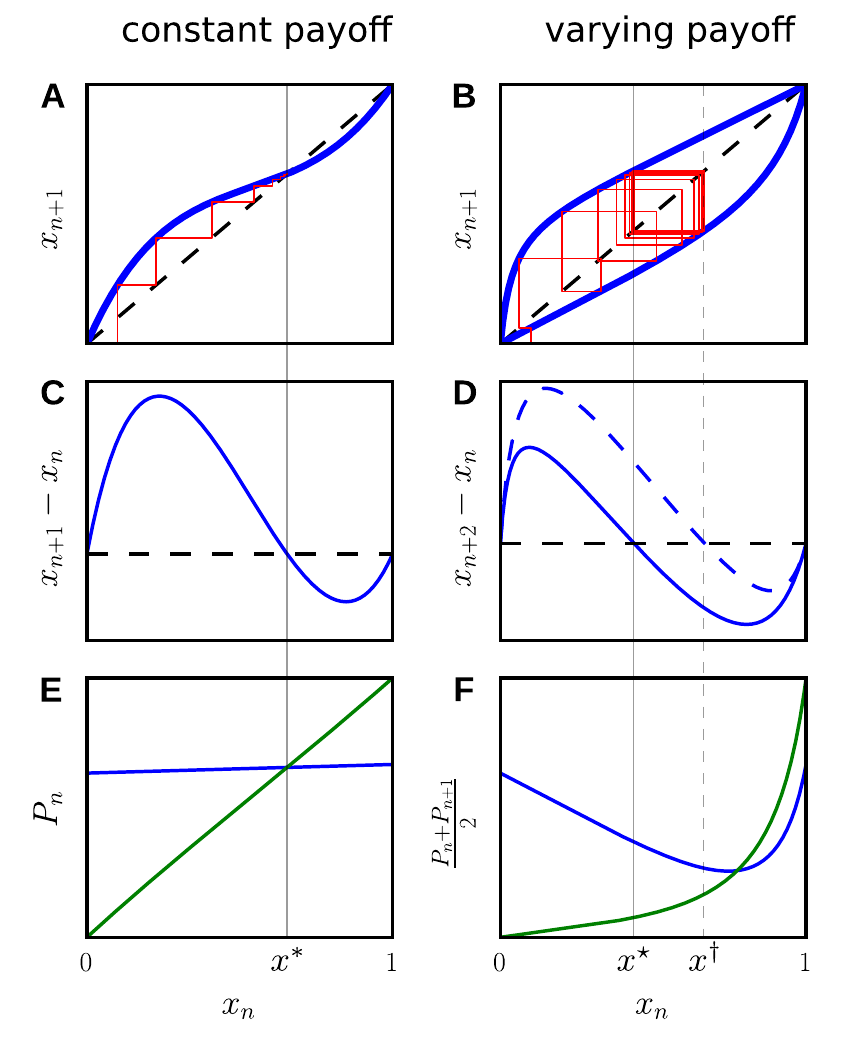}
    \caption{
    {\bf Evolutionarily stable states for constant and periodic payoff.}
    An evolutionary game with a time-constant payoff matrix (left) compared with a game with a time-varying payoff matrix (right) to exemplify the ``unfair'' stationary states.
     The former payoff matrix represents a Hawk-Dove game and the latter an alternating payoff matrix which has the same time-average as the constant Hawk-Dove game. 
            (A) and (B) show the dynamics of the two games as a Verhulst diagram with an example trajectory in red. 
            The two blue curves in (B) correspond to even and odd time points, 
            with the anomalous stationary states at $x_n^{\star}$ and $x_n^{\dagger}$.
			(C) and (D) show the difference of the state one (C) or two (D) time steps later and the current state. The zero crossings of these lines are the positions of the fixed point and the periodic points.
            (E) and (F) show the payoff of species 1 (blue) and species 2 (green).
             In (E), the equilibrium is at the same position as the fixed point. 
             In (F), species 1 receives a higher time-averaged
             payoff than species 2 at both periodic points.
            }
    \label{fig:unfair-game}
\end{figure} 

\section{Stochastic payoff fluctuations}
\label{sec:stochastic}

\subsection{Replicator equation}
How do anomalous stationary states arise from stochastic payoffs?
To avoid unnecessary technicalities, 
we consider the case of two strategies, in which the state is fully described by a scalar $x=x_1$ (because $x_2=1-x_1$) and the payoff $M=\mathbf{Y}=\begin{pmatrix} Y_{1} & Y_{2} \\ Y_{3} & Y_{4} \end{pmatrix}$ is a random matrix, where $Y_{j}$ have probability density functions $P_{Y_{j}}(y_{j})$, mean $E(Y_{j})=\mu_{j}$ and variance $\text{Var}(Y_{j})=\sigma_{j}^2$.
In short, we can write the replicator equation as
\begin{align}
X^{(t+1)} = f(X^{(t)},\mathbf{Y})
\end{align}
with $ f(x,\mathbf{y}) = x^{(t)}\cdot \frac{(y^{(t)}\mathbf{x}^{(t)})_1+b}{\mathbf{x}^{(t)T} \mathbf{y}^{(t)}\mathbf{x}^{(t)}+b} $.
In order to get a function which is injective with respect to $\mathbf{Y}$ we define a new function
\begin{align}
f'(X^{(t)},\mathbf{Y}) = \begin{pmatrix} f(X^{(t)},\mathbf{Y}) \\ Y_{2} \\ Y_{3} \\ Y_{4} \end{pmatrix}
\end{align}
This function is invertible, hence we can derive the joint probability $P_{f'}(x^{(t+1)},y_{2},y_{3},y_{4})$ from the joint probability $P_{\mathbf{Y}}(y_{1},y_{2},y_{3},y_{4}) = P(y_{1}) P(y_{2}) P(y_{3}) P(y_{4})$ by changing variables,
\begin{align}
P_{f'}(x^{(t+1)},y_{2},y_{3},y_{4}) = \left| \text{det}[ D f'^{-1} ] \right| P_{\mathbf{Y}}(f'^{-1})
\end{align}
The stochastic kernel can be derived by marginalizing over $y_{2}$, $y_{3}$ and $y_{4}$.
\begin{align}
K(x^{(t+1)}|x^{(t)}) = \int \int \int P_{f'}(x^{(t+1)},y_{2},y_{3},y_{4}) \times \nonumber \\ 
P_{Y_{2}}(y_{2}) P_{Y_{3}}(y_{3}) P_{Y_{4}}(y_{4}) \dif y_{2} \dif y_{3} \dif y_{4} 
\end{align}
The Chapman-Kolmogorov equation gives the time evolution of the probability density
\begin{align}
P^{(t)}_{X}(x) = \int_0^1 \dif x' P^{(t-1)}_{X}(x') K(x|x'),
\end{align}  
To ease the numerical evaluation we use the look-ahead-estimator \cite{Stachurski2008}
\begin{align}\label{eq:lookaheadbutnottoofar}
P^{(t)}_{X}(x) = \frac{1}{n} \sum_{l=1}^n K(x|s_l^{t-1})
\end{align}
where $\{s_l\}$ is a sample of size $n$ drawn from $P^{(t-1)}_{X}(x)$. 
Starting with an arbitrary initial distribution $P^{(1)}_{X}(x)$ and successively applying Eq.~(\ref{eq:lookaheadbutnottoofar}) converges to a stationary distribution $\rho^*(x) = P^{(\infty)}_{X}(x)$
of the stochastically driven replicator dynamics. \\
The background in Fig.~3 
shows that the stationary distributions, apart from the expected broadening, follow the behavior of the stable states derived for analogous deterministic fluctuations.

\begin{figure}[htp]
    \centering
    \includegraphics[]{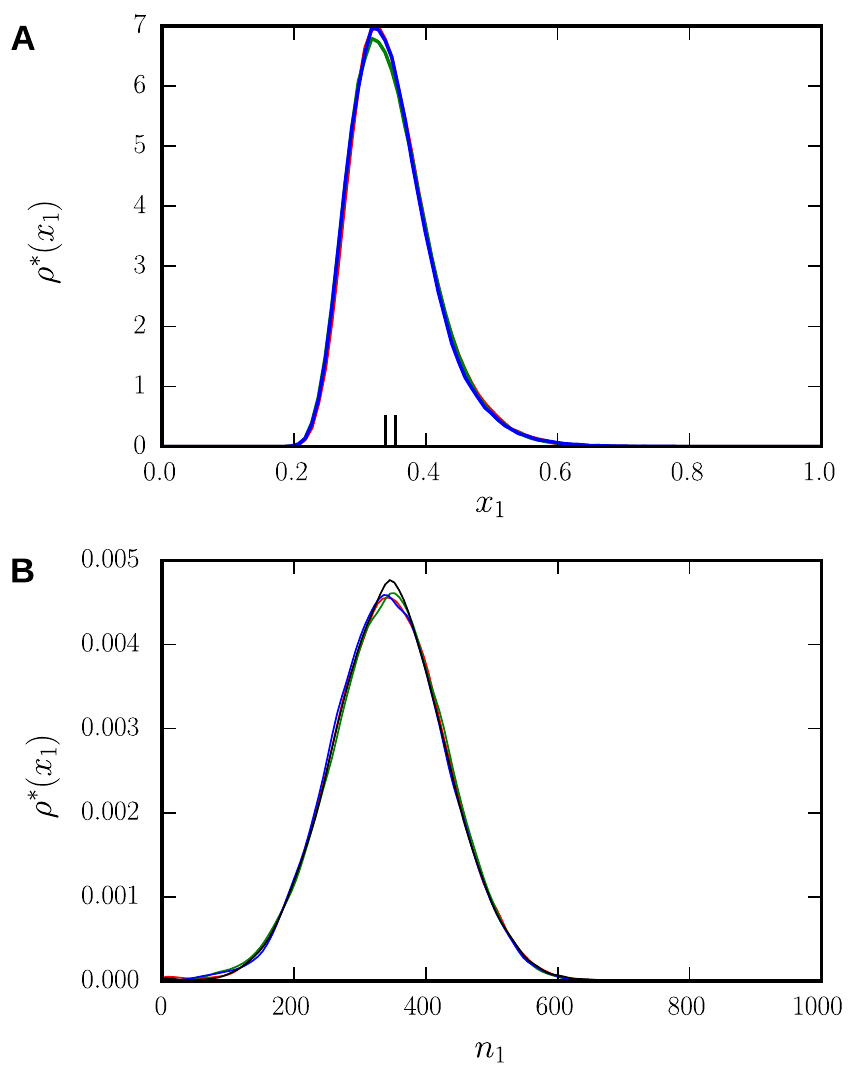}
    \caption{Stationary distributions of replicator dynamics (A) and Moran model (B) 
    for different noise sources ($\sigma=2$). Red: Uniform distributed, green: discrete distributed, blue: normal distributed, black: alternating noise.}
    \label{fig:stationary-distributions}
\end{figure}

We now demonstrate that the type of the distribution has only little effect on the stationary states.
As an example we use a game with the payoff function 
\begin{align}
M^{(t)}=\begin{pmatrix} 1 & 0.5\\ 2 & 0 \end{pmatrix}+ X \begin{pmatrix} 0 & 0\\ 1 & 0\end{pmatrix}
\end{align}
with the background fitness $b=10$.
Note that the zero-noise case of this game resembles a Hawk-Dove game. 
Fig.~\ref{fig:stationary-distributions} shows the stationary distributions $\rho^*(x_1)$ of the replicator dynamics and the Moran process, where $X$ is either a uniform, discrete, normal distributed random variable or alternations, each with variance $\sigma=2$. The higher moments of the noise distribution have little effect on the resulting stationary distribution.

\subsection{Moran processes}

\begin{figure}
    \centering
    \includegraphics[]{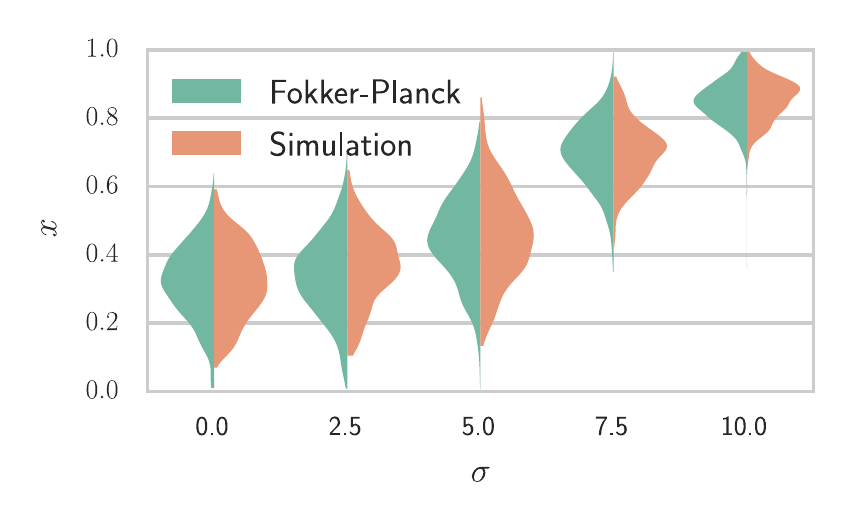}
    \caption{{\bf Anomalous stationary distributions of the Moran model: Theory and simulation.}
    Stationary distributions calculated by the Fokker-Planck equation 
    and measured stationary distributions using a simulation of the Moran model in comparison. 
    The payoff matrix is $M^{(t)} = [ 6,  5.5, 7, 5] + X^t \sigma  [0,0,1,0]$, 
    with $X^t$ randomly switching between $+1$ and $-1$.
     The population size is $N=1000$. }
    \label{fig:comparison-moran-fokkerplanck-simulation}
\end{figure}

Employment of Moran processes has been shown to be imperative for the mathematical understanding
of stochastic evolutionary game theory.
Despite being conceptionally very different from replicator dynamics, Moran processes are affected by payoff fluctuations in a similar way. \\
Consider a Moran process with population size $N$
and payoff matrix $M=\mathbf{Y}=\begin{pmatrix} Y_{1} & Y_{2} \\ Y_{3} & Y_{4} \end{pmatrix}$, where $Y_{j}$ are uncorrelated random variables with probability density functions $P_{Y_{j}}(y_{j})$ (note that also a deterministically changing payoff with period $T=2$ can be mapped to this formulation\footnote{Periodic fluctuations with period $T=2$ can be reinterpreted as uncorrelated noise: The non-zero transition probabilities are $T(i|i+2)$, $T(i|i)$ and $T(i|i-2)$ ($T(i|i)$ does not appear in the simplified master equation). With $i'=2i$ we have the same situation as with random values from a probability distribution $P(x)= \delta(x+\sigma)+\delta(x-\sigma) $.}). 
If the number of individuals playing strategy 1 is $i$, the expected payoff received by an individual playing strategy 1 or 2 is
\begin{align}
p_1(t) &= \frac{1}{N-1} \left[ y_{1}(t)(i-1) + y_{2}(t)(N-i) \right] \\
p_2(t) &= \frac{1}{N-1} \left[ y_{3}(t)i + y_{4}(N-i-1) \right]
\end{align}
With selection strength $w$ the fitness of each strategy $k=1,2$ reads
\begin{align}
f_k(t) = 1 - w + w p_k.
\end{align}
The (non-zero) transition probabilities are
\begin{align}
T(i|i+1) =  &\int \frac{f_1(\mathbf{Y}) i(N-i)}{[f_1(\mathbf{Y}) i + f_2(\mathbf{Y})(N-i)]N}  P_{\mathbf{Y}}(\mathbf{y}) \dif \mathbf{y} \nonumber \\
            &+ g(i/N) \\ 
T(i|i-1) =  &\int  \frac{f_2(\mathbf{Y}) i(N-i)}{[f_1(\mathbf{Y}) i + f_2(\mathbf{Y})(N-i)]N} P_{\mathbf{Y}}(\mathbf{y}) \dif \mathbf{y}\nonumber \\
             &+ g(1-i/N)
\end{align}
where we use the abbreviation $P_{\mathbf{Y}}(\mathbf{y}) \dif \mathbf{y}=P_{Y_{1}}(y_{1}) P_{Y_{2}}(y_{2}) P_{Y_{3}}(y_{3}) P_{Y_{4}}(y_{4}) \dif y_{1} \dif y_{2} \dif y_{3} \dif y_{4} $
and add $g(x) = \delta(x)$ to achieve reflecting boundaries\footnote{
For practical purposes (instead of a half delta function) we choose $g(x) = e^{-1000 x}$ 
which is differentiable and ensures reflecting boundaries.}.
The explicit form of the transition probabilities
 allows to 
calculate the anomalous stationary state 
as the solution of the Fokker-Planck equation for the Moran process \cite{Traulsen2006a}
\begin{align}
\partial_t \rho(x,t) = - \partial_x \left[a(x)\rho(x,t)\right]+\frac{1}{2} \partial_x^2 \left[ b^2(x)\rho(x,t)\right]
\end{align}
which reads
\begin{align}
\rho^*(x) = \mathcal{N} \exp{\left( \int_0^x \Gamma(x') \dif x' \right)} \label{eq:fokker-plack-solution}
\end{align}
for 
\begin{align}
\mathcal{N} = \int_0^1 \exp{\left( \int_0^x \Gamma(x') \dif x' \right)} \dif x,
\end{align}
and
\begin{align}
\Gamma(x) = \frac{1}{b(x)} \left(2a(x) - \frac{\dif b}{\dif x}(x) \right), 
\end{align}
where 
$a(x) = T(x|x+1/N) - T(x|x-1/N)$, 
$b(x) = \sqrt{\left( T(x|x+1/N) + T(x|x-1/N) \right)/N}$, and
            $x = \frac{i}{N}$. \\
Figure \ref{fig:comparison-moran-fokkerplanck-simulation} shows for an example how the 
stationary distributions, predicted by Eq.~(\ref{eq:fokker-plack-solution}),
change with increasing fluctuation intensities compared to stationary distributions of the simulated Moran model. 

\subsection{Correlated fluctuations}

\begin{figure}
    \centering
    \includegraphics[]{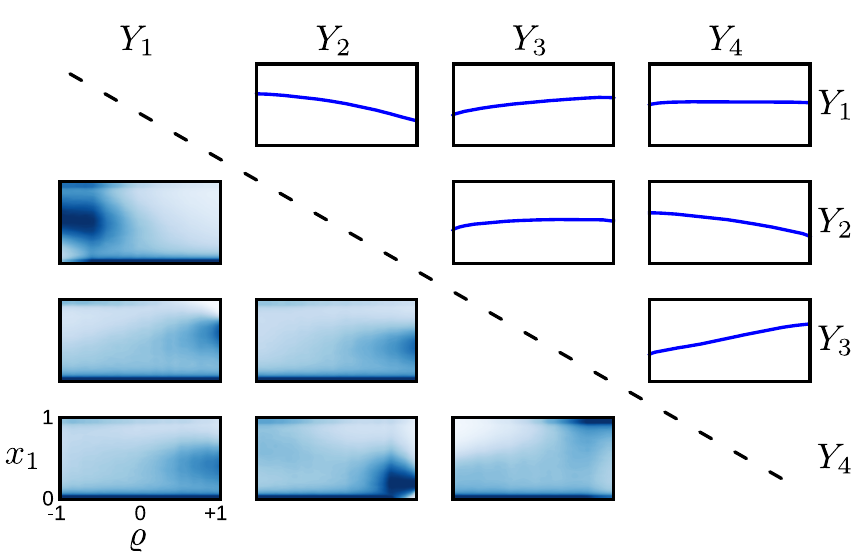}
    \caption[]{{\bf Impact of correlations.}
    Periodic points $x^\star$ (upper part)
    and stationary distributions (lower part) for fluctuations with equal intensity ($\sigma=5$) but different correlation coefficients $\varrho$.
	Each plot shows the impact of the correlation between two entries of the payoff matrix while keeping all other correlations zero.}
    \label{fig:replicator-correlation-matrix}
\end{figure}

Payoff values are not necessarily statistically independent from each other,
as for example the payoff values $M_{11}$ and $M_{12}$ of the Hawk-Dove game in Fig.~1. 
Thus, it is informative to study the effects of covariation.
Consider the general case of a game specified by the payoff matrix $M=\begin{pmatrix} Y_1 & Y_2\\ Y_3 & Y_4 \end{pmatrix}$. In order to show the impact of the correlations, we keep the intensity of the fluctuations equal and constant, $\sigma_{Y_1}=\sigma_{Y_2}=\sigma_{Y_3}=\sigma_{Y_4}=\text{const}$. 
The correlations between $Y_1$, $Y_2$, $Y_3$ and $Y_4$ are specified by six independent correlation coefficients on which the resulting stationary states depend in a nonlinear way. 
For simplicity, 
Fig.~\ref{fig:replicator-correlation-matrix} shows only the isolated 
 impact of each pairwise correlation keeping the others zero.

This shows that in addition to intensities, the anomalous stationary states are crucially determined by the correlation of the fluctuations.
Yet, we can show analytically 
that there is a special case ($\text{corr}(Y_1,Y_3)=\text{corr}(Y_2,Y_4)=1$)
for which the stationary state becomes completely independent of the fluctuation intensities.
Assume that $\mathbf{x}^*$ is the stationary state of a game with constant payoff matrix $M_0$, such that $r(\mathbf{x}^*,M_0)=1$.
If we add noise with correlation coefficient $1$ between column values,
\begin{align}
M(t)= M_0 + \tilde{M}(t) = \begin{pmatrix} a & b \\ c & d \end{pmatrix} + \begin{pmatrix} f_1(t) & f_2(t) \\ f_1(t) & f_2(t) \end{pmatrix} ,
\end{align}
then
\begin{align}
r\left( \mathbf{x}^*,M_0+\tilde{M}(t) \right) = \frac{(M_0 \mathbf{x}^*)_i + (\tilde{M}(t)\mathbf{x}^*)_i + b}{ \mathbf{x}^{*T} M_0 \mathbf{x}^* + \mathbf{x}^{*T} \tilde{M}(t) \mathbf{x}^* + b} \\
                                                 = \frac{(M_0 \mathbf{x}^*)_i + (f_1(t) x_1^* + f_2(t) x_2^*) + b}{ \mathbf{x}^{*T} M_0 \mathbf{x}^* + (f_1(t)x_1^{*} + f_2(t)x_2^{*}) + b} = 1
\end{align}
where the second step uses $x_1^2+x_1x_2=x_1$ and $x_2^2+x_1x_2=x_2$ 
and the last step uses $(M_0 \mathbf{x}^*)_i = \mathbf{x}^{*T} M_0 \mathbf{x}^*$ following from the assumption.
Consequently, in this case the stationary state does not depend on the noise intensity.

\section{Classification of games with payoff fluctuations}
\label{sec:classification}

\subsection{Generalized criteria}
\begin{figure}
    \centering
    \includegraphics[]{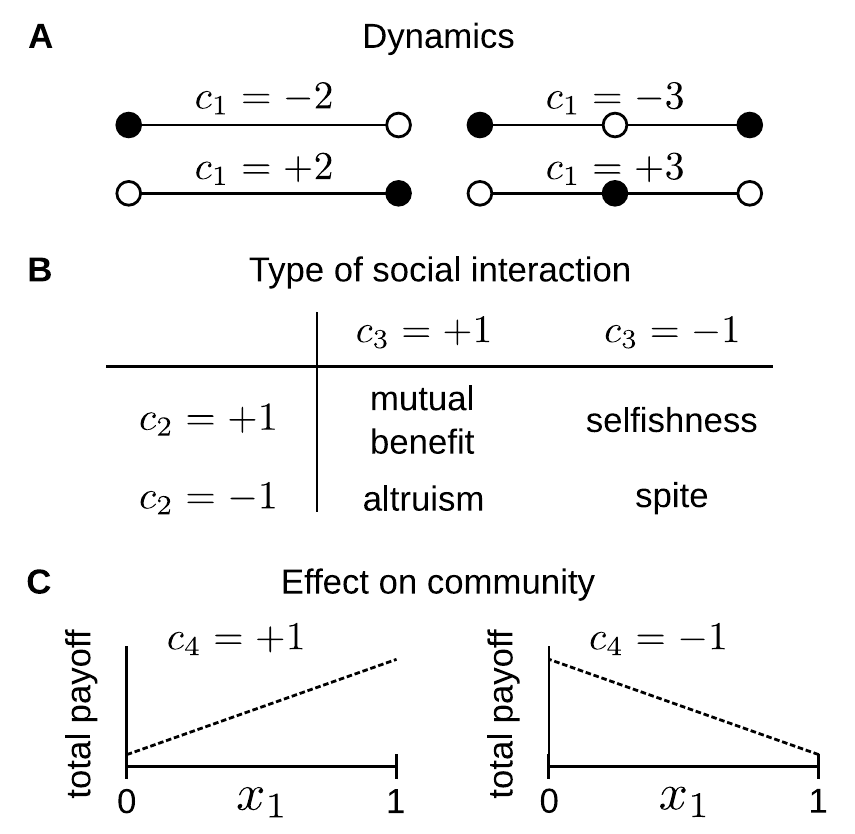}
    \caption[]{{\bf Case differentiation of game characteristics.}
    An evolutionary game with two strategies and constant payoff has one out of
(A) four possible dynamical structures (either two or three stationary states with the first one either stable or unstable),
(B) four possible combinations of strategy 1's impact on the payoff of the individuals (positive or negative impact on the payoff of strategy 1 and 2 players), 
and (C) two possible kinds of strategy 1's impact on the total payoff of all players (positive or negative).
    }
    \label{fig:case-differentiation}
\end{figure}

A symmetric game defined by a constant $2\times 2$ payoff matrix can be classified as one out of 12 game classes 
with distinct dynamical structures, e.\,g. Prisoner's Dilemma, Hawk-Dove game, etc. 
This traditional classification is based on the rank of the four values in the payoff matrix, see middle column in table \ref{tab:list-of-games}.
The name of a game allows a more intuitive understanding than
the position in the four dimensional payoff space. 
However, this classification cannot be applied to time-varying payoff matrices because the ranks 
may be time-dependent. 
Therefore we propose a classification for evolutionary games based on three characteristics: 
(1) the dynamics of the evolutionary game (the number of stationary states and their stability), 
(2) the type of social interaction (how the payoff differs between stationary states for one player compared to the other player) and 
(3) the effect on the community (how the total payoff of player one and two differs between stationary states). 
The classification scheme and its criteria are summarized 
in Fig.~\ref{fig:case-differentiation}.
Based on these criteria a game class is defined as a tuple $ [c_1,c_2,c_3,c_4] $, where
\begin{align}
\begin{split}\label{eq:cs}
c_1 &= \text{Sign}\left(\frac{\dif u_1}{\dif x_1}(0)\right)\cdot n^* \\
c_2 &= \text{Sign}\left( P_1(1)-P_1(0) \right) \\
c_3 &= \text{Sign}\left( P_2(1)-P_2(0) \right) \\
c_4 &= \text{Sign}\left( \langle P \rangle(1) - \langle P \rangle(0) \right),
\end{split}
\end{align}
where $u_1=\frac{\dif x_1}{\dif t}$, $P_i(x) = \left( M \begin{pmatrix} x\\ 1-x \end{pmatrix}\right)_i$ denotes the payoff of a strategy $i$ player, $\langle P\rangle=x P_1+(1-x) P_2$ 
the average payoff in the population and $n^* = \|\{ x_1^*: u_1(x_1^*)=0\}\|$ the number of anomalous stationary states.
This classification can be applied to games with varying payoff matrices and even games with nonlinear payoff functions. 
The scheme is developed for time-continuous dynamics. 
The formulation for time-discrete dynamics is analogous.
Note also that the criteria (c2-c4) of Eqs.\ (\ref{eq:cs})
can be written 
in a more general form to describe also non-monotonic payoff functions.\\
Table \ref{tab:list-of-games} lists the 12 traditional games defined by the payoff rank criteria and their corresponding definitions with the presented generalized criteria. 

\subsection{Proof that payoff rank criteria and generalized criteria are equivalent in case of constant payoffs}
\label{sec:proof-equivalence-constant-payoff}

Since the method is the same for all games we show it only for the Prisoner's Dilemma to exemplify the proof. 
We assume that the dynamics of the game are described by the continuous replicator equation 
$u_1:= \dot{x}_1 = x_1 ( (M\mathbf{x})_1 - \mathbf{x}^T M\mathbf{x})$
with a constant payoff matrix $M=\begin{pmatrix} m_{1} & m_{2} \\ m_{3} & m_{4} \end{pmatrix}$. 
According to the generalized criteria a Prisoner's Dilemma is defined as $[-2, +1, +1, +1]$. The $-2$ tells us that the first stationary state at $x_1=0$ is stable and the second at $x_1=1$ is unstable, 
\begin{align}
\frac{\dif {u}_1}{\dif x_1}(0) < 0 \Leftrightarrow m_2 < m_4 \label{eq:condition1} \\
\frac{\dif {u}_1}{\dif x_1}(1)> 0 \Leftrightarrow m_3 > m_1 \label{eq:condition2}.
\end{align}
Further the three $+1$ tell us that the payoff of both players and the total payoff of the population at $x_1=1$ is higher than at $x_1=0$,
\begin{align}
 P_1(1) > P_1(0) \Leftrightarrow m_1 > m_2 \label{eq:condition3}\\
 P_2(1) > P_2(0) \Leftrightarrow m_3 > m_4 \label{eq:condition4}\\
 \langle P \rangle (1) > \langle P \rangle (0) \Leftrightarrow m_1 > m_4. \label{eq:condition5} 
\end{align} 
Criteria~(\ref{eq:condition1}) to (\ref{eq:condition5}) are equivalent to $m_2<m_4<m_1<m_3$ or in the payoff rank notation $[3,1,4,2]$, which defines a traditional Prisoner's Dilemma.

\subsection{Application of the generalized criteria on an alternating payoff matrix}
As an illustrative example we show that the game in Fig.~2 
at noise intensity $8$, where the payoff matrix is $M^{(t)}=\begin{pmatrix} 1.1 & 0.8 \\ 2 & 0 \end{pmatrix} + 8\cdot(-1)^t \begin{pmatrix} -0.33 & 1 \\ 0 & 0\end{pmatrix}$, is a Prisoner's Dilemma. \\ 
As we can see in the figure there are two stationary states, a stable state at $x_1=0$ and an unstable state at $x_1=1$, consequently $c_1 = -2$.
From the expected payoff $P_i(x_1) = \frac{1}{2} \left( M^\text{even t}\begin{pmatrix} x_1 \\ 1-x_1 \end{pmatrix} + M^\text{odd t}\begin{pmatrix} x_1 \\ 1-x_1 \end{pmatrix} \right)$ evaluated at the stationary states ($P_1(0)=0.8$, $P_1(1)=1.1$, $P_2(0)=0$ and $P_2(1)=2$) it follows that $c_2=+1$ and $c_3=+1$. For the last criteria we evaluate the population payoff $\langle P \rangle(x_1)=x_1 P_1 + (1-x_1)P_2$ at the stationary states ($\langle P \rangle(0) = P_2(0)$ and $\langle P \rangle(1) = P_1(1)$), which results in $c_4=+1$. 
To summarize, the game satisfies the generalized criteria $[-2,+1,+1,+1]$. According to table~\ref{tab:list-of-games} this defines a Prisoner's Dilemma.

\begin{table}
\centering
\begin{tabular}{ c|c|c }
\textbf{Name} & \thead{\textbf{Payoff rank}\\ \textbf{criteria}}  & \thead{\textbf{Generalized}\\\textbf{criteria}} \\ \hline \hline
Hawk-Dove            & \makecell{$[3,2,4,1]$ \\ $[1,4,2,3]$} & \makecell{$[+3,+1,+1,+1]$ \\ $[+3,-1,-1,-1]$} \\ \hline 
Battle             & \makecell{$[2,3,4,1]$ \\ $[1,4,3,2]$} & \multirow{2}{*}{\makecell{ $[+3,-1,+1,+1]$ \\ $[+3,-1,+1,-1]$} } \\ \cline{1-2} 
Hero               & \makecell{$[1,3,4,2]$ \\ $[2,4,3,1]$} & \\ \hline
Compromise         & \makecell{$[1,2,4,3]$ \\ $[3,4,2,1]$} & \makecell{$[-2,-1,+1,-1]$ \\ $[+2,-1,+1,+1]$} \\ \hline 
Deadlock           & \makecell{$[2,1,4,3]$ \\ $[3,4,1,2]$} & \makecell{$[-2,+1,+1,-1]$ \\ $[+2,-1,-1,+1]$} \\ \hline 
Prisoner's Dilemma & \makecell{$[3,1,4,2]$ \\ $[2,4,1,3]$} & \makecell{$[-2,+1,+1,+1]$ \\ $[+2,-1,-1,-1]$} \\ \hline
Stag Hunt          & \makecell{$[4,1,3,2]$ \\ $[2,3,1,4]$} & \makecell{$[-3,+1,+1,+1]$ \\ $[-3,-1,-1,-1]$} \\ \hline
Assurance          & \makecell{$[4,1,2,3]$ \\ $[3,2,1,4]$} & \multirow{2}{*}{ \makecell{$[-3,+1,-1,+1]$ \\ $[-3,+1,-1,-1]$} } \\ \cline{1-2}
Coordination       & \makecell{$[4,2,1,3]$ \\ $[3,1,2,4]$} & \\ \hline
Peace              & \makecell{$[4,3,1,2]$ \\ $[2,1,3,4]$} & \makecell{$[+2,+1,-1,+1]$ \\ $[-2,+1,-1,-1]$} \\ \hline
Harmony            & \makecell{$[4,3,2,1]$ \\ $[1,2,3,4]$} & \multirow{2}{*}{ \makecell{$[+2,+1,+1,+1]$ \\ $[-2,-1,-1,-1]$} } \\ \cline{1-2}
Concord            & \makecell{$[4,2,3,1]$ \\ $[1,3,2,4]$} & \\ \hline 
\end{tabular}
\caption{\textbf{Criteria for strict symmetric games.} The middle column shows the rank of the values in the payoff matrix $M=[a,b,c,d]$, e.\,g. $[3,2,4,1]$ means $d<b<a<c$ \cite{Bruns2015}. The right column shows the values of the criteria defined in the text. }
\label{tab:list-of-games}
\end{table}

\end{document}